\documentstyle[aps,pre]{revtex}
\makeatother
\def\abstract{\par
\ifpreprintsty %
\vskip2.5pc
\begin{center}%
{\large \abstractname\par}%
\end{center}%
\vskip.5pc
\fi
\bgroup
\ifdim\prevdepth=-1000pt \prevdepth0pt\fi
\hsize\columnwidth
\hsize2\columnwidth%<<ADDED>>
\if@twocolumn\else\leftskip=0.10753\textwidth \rightskip\leftskip\fi
\dimen0=-\prevdepth \advance\dimen0 by17.5pt \nointerlineskip
\small\vrule width 0pt height\dimen0 \relax
}
\def\endabstract{\par\egroup}
\makeatletter

\begin{document}

\title{
Collective motion occurs inevitably
in a class of populations of globally coupled chaotic elements
}
\author{Naoko NAKAGAWA}
\address{
FRP, Laboratory for Information Synthesis, 
The Institute of Physical and Chemical Research (RIKEN),
2-1 Hirosawa, Wako, Saitama 351-01, Japan}

\author{Teruhisa S. KOMATSU}
\address{
Department of Pure and Applied Sciences, 
College of Arts and Sciences,
University of Tokyo, 
Tokyo 153, Japan}

\twocolumn[
\maketitle

\begin{abstract}

We discovered numerically a scaling law obeyed by the amplitude of collective motion in large populations of chaotic elements.
Our analysis strongly suggests that such populations generically exhibit
collective motion in the presence of interaction, however weak it may be.
A phase diagram for the collective motion,
which is characterized by peculiar structures similar to 
Arnold tongues, is obtained.

\end{abstract}
\ \\
]

\section{Introduction}

Systems consisting of large populations of interacting dynamical elements 
are widely distributed in nature
from communities of ants to biological cell assemblies and neural networks.
Many such populations are similar in the respect that they exhibit 
various kinds of collective behavior.
One possible approach to the mathematical study of such collective behavior 
is to concentrate on certain idealized models 
such as interacting limit cycles and chaotic maps.
An abundance of literature has been devoted to the study of these models
\cite{Winfree,Kuramoto,oscillator1,Kaneko_CML,Kaneko_GCM,Chate_Manneville,Perez_Cerdeira,Pikovsky_Kurths,oscillator2}.
It is suggested in some foregoing studies 
\cite{Chate_Manneville,Perez_Cerdeira,Pikovsky_Kurths,oscillator2} that
a population of this type as a whole exhibits low dimensional behavior.
This seems to be true even when the individual elements appears to be
mutually uncorrelated,
a situation in which it would seem that the population could only be 
regarded as a dynamical system of 
extremely high dimension.
It still remains unclear whether such collective behavior could be understood 
in terms of low-dimensional dynamical systems.

Without directly attacking the above stated problem,
we take in the present Paper a slightly different approach
in which we focus on a non-specific property of collective motion,
a long-time average of the amplitude of collective motion.
As we find, this approach, with which we intentionally avoid addressing
specific features of collective motion, turns out to be extremely useful.
As model systems, we have chosen globally coupled tent maps, 
because these systems are particularly well suited 
for detailed numerical analysis.
Using these maps, we have discovered a scaling law characterizing 
the amplitude of collective motion
which holds for a particular series of parameter values.
Further, we argue on the basis of numerical evidence 
that this type of collective motion should occur generically
if an interaction exists, no matter how weak this interaction may be.
Finally, a phase diagram for the collective motion is obtained,
and the existence in it of peculiar structures similar to Arnold tongues 
with various scales is confirmed.

%%%%%%%%%%%%% MODEL %%%%%%%%%%%%%%%
\section{Model : Globally coupled Tent maps}

Globally coupled maps (GCM) are given by an assembly of $N$ elements 
whose behavior is determined by $N$ identical maps with all-to-all coupling. 
The individual elements are then under the influence of a common 
internal field which may be referred to as a mean field \cite{Kaneko_GCM}.
We assume that a single isolated element 
evolves according to $X_{n+1}=f(X_n)$,
where $n$ designates discrete time steps.
Under the interaction through the mean field $h_n$,
the $i$-th element is then assumed to evolve as:
\begin{equation}
X_{n+1}^{(i)}=f(X_n^{(i)})+K h_n, 
\label{eqn:gtent}
\end{equation}
where $K$ is the coupling strength.
In this Paper, we consider the situation in which $f$ is a tent map:
\begin{equation}
f(X)=-a|X|+\displaystyle{{a-1}\over 2},
\label{eqn:tent}
\end{equation}
and the mean field $h_n$ is defined as:
\begin{equation}
h_n \equiv {1\over N}\sum\limits_{j=1}^Nf(X_n^{(j)}).
\end{equation}
Thus, our system is characterized by the two parameters $a$ and $K$
in addition to the total number of elements $N$.
Each tent map has a band splitting point $a=\sqrt{2}$
and we thus stipulate that $a$ satisfies $\sqrt{2}<a<2$.
It is thereby ensured that the population will never split 
into sub-populations.
If the system size $N$ is a finite, 
this finiteness becomes the source of fluctuations of the mean field.
Such an effect may obscure pure collective motion.
Thus we work with the limit of large $N$.
In fact, we confirmed that finite size effects can be regarded as noise
acting on the pure collective motion.
For $N\rightarrow \infty$, the population dynamics of GCM 
can be described by the Frobenius-Perron equation \cite{Kaneko3}
for the distribution $\rho (X)$:
\begin{eqnarray}
&&\rho_{n+1}(X)=\int\delta(X-f(X')-Kh_n)\rho_n(X')dX',
\label{eqn:FP}\\
&&h_n=\int f(X')\rho_n(X')dX'.
\label{eqn:op}
\end{eqnarray}
We worked out a numerical scheme for the exact integration of
Eqs.(\ref{eqn:FP}) and (\ref{eqn:op}) whose precision is limited only 
by round-off errors.
Our scheme is almost the same as that proposed by Morita \cite{Morita}, 
although these numerical schemes were developed independently.
We have confirmed that collective motion appearing in the system 
described by Eqs.(\ref{eqn:FP}) and (\ref{eqn:op}) is almost independent of
the initial distribution.
For this reason we present numerical results in this Paper
were obtained using only a single initial distribution, 
uniform over the interval $[f\circ f(0):f(0)]$.

\section{Collective motion in globally coupled tent maps}

In some previous works \cite{Perez_Cerdeira,Kaneko3},
it has been argued that the model constituted by Eqs.(\ref{eqn:gtent}) and 
(\ref{eqn:tent}) does not exhibit collective motion.
However, our careful analysis yields a contradictory conclusion.
Collective motion can be observed through the dynamics of the order 
parameter $h_n$ given by Eq.(\ref{eqn:op}).
Figures \ref{fig:snapshot}(a) and \ref{fig:snapshot}(b) are return maps 
of $h_n$. 
There we see that
the fluctuation of $h_n$ undergoes quasi-periodic motion for small $K$,
but for larger $K$ it displays more complicated motion 
possibly with fine structure.
Roughly speaking, larger $K$ values results in more complicated 
collective behavior.

The observed collective motion is similar to that found in other models 
\cite{Chate_Manneville,Perez_Cerdeira,Pikovsky_Kurths},
except that the scale of the collective motion for the present model 
is much smaller.
Because of this smallness, 
the previous works \cite{Perez_Cerdeira,Kaneko3} have failed to detect
the collective motion for the present model.
Based on their conclusion that collective motion does not exist here, 
they conjectured that the presence of window structures for the elements 
(which tent maps never possess)
are necessary for the occurrence of collective motion,
but our results shows this is not the case.

%%%%%%%%% hill %%%%%%%%%%%%%
\section{Parameter dependence of the collective motion : hilly structure}

We concentrate on the amplitude $F$ of the collective motion without going 
into any detailed structure of the dynamics.
We have
\begin{equation}
F\equiv \sqrt{<(h_n-<h_n>)^2>},
\label{eqn:F}
\end{equation}
where $< >$ represents a long-time average.
In Fig.\ref{fig:a_F}, the dependence of $F$ on $a$ is shown 
for two values of $K$.
The horizontal axis represents $m\equiv -\log_a(2-a)$,
which is a monotonically increasing function of $a$ on (0,2),
tending to $\infty$ as $a\rightarrow 2$.
The figure shows the range of many hills of $F$,
which we call `hilly structure'.
As the value of $K$ is decreased, the number of the hills becomes larger,
while the average size of each hill becomes smaller.
For larger $a$ (i.e. larger $m$), the hilly structure is difficult 
to discern in this figure,
but its persistence can be confirmed by magnifying the scale.
Hilly structure seems to exist for arbitrarily small values of $K$.
Furthermore, based on this figure we expect that 
for a given value of $K$, collective motion exists for almost all $a$.
These points will be discussed in further detail below.

%%%%%%%%%%% golden values %%%%%%%%%%%%%%%
\section{Golden and silver values of parameter "\lowercase{$a$}"}

We find that the role of a particular series of parameter values of 
$a$ \cite{Mori}, 
which we call `golden values',
play a crucial role in the formation of such hilly structure.
We find in Section VII that these parameter values are situated 
in the middle of each hill for sufficiently small values of $K$.
Some examples are indicated in Fig.\ref{fig:a_F} with the notation
$g_3 \sim g_7$.
When $a$ is identical to one of the golden values, 
one isolated tent map possesses the property that
a trajectory beginning at the peak C of the tent map returns to C
after $p$ steps.
We denote a golden value defined in this way as $g_p$.
Incidentally, we use the term `golden value' because the value $g_3$ equals 
the golden mean.

We refer to another special series of parameter values of $a$ as
`silver values'.
These are associated with the valleys between neighboring hills,
as argued in Section VIII.
For instance, the silver values $s_4 \sim s_7$ are indicated 
in Fig.\ref{fig:a_F}.
When $a$ is identical to $s_p$, 
a trajectory beginning at C falls into a fixed point of the tent map 
after $p$ steps.
Golden and silver values exist densely if we allow $p$ to take all natural 
number values,
while they occupy only a vanishing measure on the line of $a$.
We note that $g_p$ and $s_p$ are generally not unique 
because of the multiplicity of $p$-periodic orbits,
whose number increases almost exponentially with $p$.
Despite this non-uniqueness, $g_p$ ($s_p$) is used below 
to represent a single golden (silver) value.

\section{Scaling law of Collective motion at golden values}

We now consider the collective motion as a function of the coupling strength 
$K$, with $a$ confined to golden values.
We found numerically that the amplitude $F$ of the collective motion
obeys the law
\begin{equation}
KF\sim e^{-{\alpha\over K}},
\label{eqn:exp_K}
\end{equation}
where $\alpha$ is a positive constant depending on 
the golden values in question.
For any golden value,
the above scaling form seems to hold,
as long as $K$ is sufficiently small. 
In Fig.{\ref{fig:gn0_exp}}, 
the relation between $KF$ and $K^{-1}$ is displayed for those golden values
which are situated in the middle of some representative hills 
in Fig.\ref{fig:a_F}.
The formula in Eq.(\ref{eqn:exp_K}) reveals some important properties 
of the collective motion.
First, the collective motion persists even as $K\rightarrow 0$.
Second, the observed quasi-periodic motion does not appear through 
the conventional route of Hopf bifurcation.
This follows from the fact that if the quasi-periodicity were 
due to a Hopf bifurcation,
then we would find that $F\sim \sqrt{|K-K_c|}$ above a bifurcation point $K_c$.
This is clearly in contradiction with Eq.(\ref{eqn:exp_K}).
In our study we could not even find an indication of a bifurcation.
Third, the simple relation in Eq.(\ref{eqn:exp_K}) survives 
even when the collective motion becomes more complicated.
Note that different modes of collective motion as illustrated in 
Figs.\ref{fig:snapshot}(a) and (b), lie on a common line of 
Fig.{\ref{fig:gn0_exp}}.

%%%%%%%%%%% hill formation %%%%%%%%%%%%%
\section{Formation of hilly structure}

The appearance of hilly structure, seen in Fig.\ref{fig:a_F},
can be understood from two arguments, concerning hills and valleys.
The following discussion regarding hills based on Eq.(\ref{eqn:exp_K}) 
partially explains how 
collective motion comes to characterize almost all values of $a$,
not just the golden values.

At a given value of $K$ each hill of $F$ contains a number of golden values.
One of these golden values is found to be representative 
of this hill. 
We call this a `key golden value'.
In order to define key golden values, 
we studied the growth of the width of hills with the decrease of $K$.
Figure \ref{fig:width} shows how the width $W$ of the hill 
around a given golden value changes with $K^{-1}$.
The width $W$ around a given $g_p$ is defined by $W=|a_L-a_R|$,
where $a_L$ and $a_R$ are the two values of $a$ on either side of $g_p$ 
such that the amplitude $F$ at $a=a_L$ and $a=a_R$ is half of $F(g_p)$.
The golden values chosen here are the same ones as 
in Figs.\ref{fig:a_F} and \ref{fig:gn0_exp}.
In this figure, we see that $W$ behaves as a function of $K$ similar to 
that in Eq.(\ref{eqn:exp_K}).
Actually, the width $W$ for a given $g_p$ is proportional 
to $KF$ calculated for the same $g_p$, as long as $K$ is sufficiently small.
This implies that the hill narrows down to a point, corresponding to $a=g_p$.
Based on this relation, we define `key golden values' of a given hill 
which exists at a given $K_0$ as follows:
With the decrease of $K$ from $K_0$,
the width of a hill narrows down to a single
point according to the growth law for $W$.
The value of $a$ for this single point is 
that of the key golden value of this hill.
The definition of key golden values will be refined as
golden values displaying the scaling law Eq.(\ref{eqn:exp_K}) 
for values of $K$ smaller than $K_0$.

The formation of the hills may roughly be understood 
from the following argument.
At a golden value $g_p$, the invariant measure for a single map, 
 i.e. $\rho(X)$ at the equilibrium solution of the Frobenius-Perron equation
with $K=0$,  has the rather simple form \cite{Mori} 
of a ($p-2$)-step function, as displayed in Fig.\ref{fig:distribution}(a).
As $a$ deviates slightly from $g_p$, the change of the invariant measure 
remains small, although there may appear some fine structure.
When a weak interaction is switched on, 
the solution $\rho_n(X)$ of Eqs.(\ref{eqn:FP}) and (\ref{eqn:op}) with $a=g_p$ 
continues to still assume a shape quite similar to the invariant measure 
$\rho(X)$  mentioned above, 
but it is accompanied by intermittent peaks 
(see Fig.\ref{fig:distribution}(b)) 
whose widths are roughly given by $KF$.
The situation remains almost the same 
when $a$ differs slightly from $g_p$.
If the effect of these peaks on the amplitude of the collective motion is 
more important than the change in the invariant measure 
caused by a small deviation of $a$ from $g_p$,
the statistical average $F$ will be insensitive to the value of $a$ 
around $g_p$.
If the deviation of $a$ from $g_p$ becomes too large, 
however, the corresponding $F$ will be dominated 
by the influence of other golden values corresponding to other hills.

Hills formed in the manner described above are consistent with
hills obeying the growth relation of the width $W$ (see Fig.\ref{fig:width}).
Further, the above reasoning applies to every golden value.
Thus every golden value can be considered as a key golden value 
around which its own hill is formed. 
This conclusion may appear to be inconsistent with the fact 
that in a given hill there are an infinite number of golden values.
Actually, 
for a given $K$, most golden values in a given hill are {\it not}
key golden values.
We have found that around such a $K$ value, 
the amplitude $F$ for these golden values does not obey 
the scaling law of $F$.
This will be discussed in Section IX.
It is now clear that collective motion is not confined only to 
values corresponding to golden values. 
It occurs over the intervals 
corresponding to hills existing around these golden values.

%%%%%%%%%%%%% valley %%%%%%%%%%%%%
\section{Collective motion is dominant : property around valleys}

The hilly structure seen in Fig.\ref{fig:a_F} can be understood 
by analyzing the structure of $F$ in the valleys between two neighboring hills.
From this analysis we are able to conclude that
collective motion occurs generically in the $a$-$K$ plane.
We specifically investigate the parameter dependence of $F$ in valleys.
As is seen from Fig.\ref{fig:a_F}, 
we find that some silver values lie precisely at the minimum points of $F$
in the valleys.
Figure \ref{fig:s1_exp} shows that 
when the value of $a$ is near a silver value $s_p$, 
we obtain the following relation 
for sufficiently small values of $K$:
\begin{equation}
F\propto |a-s_p|^{\beta}  \qquad (\beta\simeq 1).
\label{eqn:s_p}
\end{equation}
It is clear from this relation that collective motion disappears 
only at the point $a=s_p$ \cite{Chawanya}.
This relation seems to hold for any silver value
when the value of $K$ is sufficiently small.
This fact strongly suggest that the total measure of values of $a$ at which 
collective motion does not exists is vanishingly small,
because the total measure corresponding to silver values is vanishing,
although they exist densely on the line of $a$.
The remaining values of $a$ are characterized by hills which extend from
key golden values with widths which increase with $K$.
Hence the occurrence of the collective motion is dominant 
over the parameter space of $a$.
Because of the scaling law existing at golden values,
this situation holds even for vanishingly small values of $K$.

%%%%%%%%%%%% merging %%%%%%%%%%%%%%%
\section{Merging of hills}

Figure \ref{fig:s1_exp} also indicates that the valleys disappear
for sufficiently large values of $K$.
For each silver value,
we find a critical value $K_c$ for the disappearance of 
the corresponding valley.
When $K$ exceeds such a $K_c$, a valley which existed 
at a certain silver value up to $K=K_c$ suddenly disappears.
This occurs because two hills lying on either side of this silver value
merged into one at $K=K_c$.
As a result, the amplitude $F$ at this value of $a$ exhibits a sudden increase.
As described above, each hill possesses a key golden value.
After the merging of two hills, one of the two key golden values 
characterizing the previously existing hills, say $g_{p_1}$, 
ceases to be a key golden value, while the other persists.
As a result the $K$-dependence of $F$ at $a=g_{p_1}$ changes, 
and typically comes to behave as illustrated in Fig.\ref{fig:merge}.
We see in this figure that the $K$-dependence of $F(g_4)$ is maintained,
while that of $F(g_{17})$ changes 
above $K_c$, where it comes to exhibit $K$-dependence similar to $F(g_4)$.

The merging of two hills leads to a sudden expansion of the width of 
one of the hills, accompanied by the disappearance of the other.
If a hill associated with some key golden value $g_p$ successively 
absorbs the neighboring hills with the increase of $K$,
then the width $W$ of this hill also grows successively.
On the other hand, we have already seen in Fig.\ref{fig:width} 
a growth law for $W$.
These observations suggest that the values of $K_c$ characterizing
the disappearance of valleys around silver values should be 
correlated with our growth law for $W$.
In the case of Fig.\ref{fig:valley_disappear}, 
the golden value $g_4$ persists as a key golden value
over a wide range of $K$.
We calculated the values of $K_c$ for a series of silver values.
Each of these silver values was chosen in such a way that 
in a certain range of $K$, it lies at the minimum point between the hill 
associated with $g_4$ and a neighboring hill.
In fact, there should exist many more silver values of this kind 
for a given $g_4$ than those displayed in this figure.
We see from this figure that the value of $K_c$ for a given silver value
depends on the distance of this silver value from $g_4$,
and that this dependence is consistent with the growth law for $W$
seen in Fig.\ref{fig:width}.
In many cases, merging occurs between hills of vastly different sizes.
In this case, the collective motion characterizing the smaller hill 
seems to be replaced suddenly by that for the larger one.
In such cases, the hill width may appear to grow almost continuously.

From our study presented so far, 
the hilly structure of the amplitude $F$ of the collective motion
can be understood.
This structure consists of a series of hills and valleys.
Only at a minimum point of $F$ in each valley, collective motion disappears.
Thus the hilly structure implies that under a given coupling strength $K$
the collective motion occurs inevitably for almost all values of $a$.
Furthermore, the following properties for the collective motion 
are now clear:
There is a scaling law for the amplitude $F$, 
a growth law for the hill width $W$,
and a successive merging of hills resulting in a decrease 
in the number of the hills and hence the number of key golden values.
Clearly these features are interrelated.

%%%%%%%%%%% diagram %%%%%%%%%%%%%%
\section{Phase diagram}

We now consider a phase diagram of collective motion in the $a$-$K^{-1}$ plane
from the viewpoint of the amplitude $F$.
For each golden value $g_p$, we define a phase as the region of a hill 
(defined by a values of $F$ greater than some threshold value $F_{th}$) 
associated with a certain $g_p$.
Here, the definition is meaningful only if $g_p$ represents 
a key golden value.
Since golden values exist densely, such a diagram will be split into
an infinite number of domains as $K\rightarrow 0$,
each reducing to a single point corresponding to a golden value.
The extension of each domain becomes larger with the increase of $K$,
as implied from Eq.(\ref{eqn:exp_K}),
and this necessarily results in the successive merging of domains,
leaving only a few for sufficiently large $K$.
Thus our phase diagram resembles that for the phase locking between 
oscillators.
The latter is also characterized by the merging and splitting of 
domains, in this case synchronized domains, referred to as Arnold tongues.
We note, however, there is one property of our phase diagram 
which distinguishes it from that in the case of phase locking.
In the present case, in the $F_{th} \rightarrow 0$ limit, 
the domains of collective motion comprise all but a measure $0$ set of values
of $a$,
even for vanishingly weak coupling.

In Fig.\ref{fig:diagram}, we show a blowup of 
the phase diagram which contains only those domains of collective motion
for $g_p$ with $p\leq 11$.
The numerical procedure for obtaining this picture essentially follows
that for Fig.\ref{fig:width}.
We determined the domain for each $g_p$ with a given $K$ first,
displaying it as a painted region of $a$.
We then did the same for various $K$, thereby producing Fig.\ref{fig:diagram}.
Note that the domains separated for a given $K$ correspond to
distinct hills with distinct $g_p$. 
As expected, a pair of neighboring domains merge at a certain value of $K$
at which a valley disappears.
Thus, the number of separated domains decreases with the increase of $K$.
After several merging events, 
we are left with only one painted domain associated with $g_4$.

%%%%%%%%%%%%% universality %%%%%%%%%%%%%%%
\section{Discussion : Any universality class?}

The foregoing arguments suggest that collective motion occurs 
generically for arbitrarily weak coupling.
Such a result seems to be closely related with 
some global properties of the individual maps 
such as the topological arrangement of golden values.
It is our conjecture that the same conclusion  holds 
for such globally coupled unimodal maps satisfying $|f'(X)|\geq 1$ for any $X$.
Although not reported in the present Paper, 
we also confirmed that similar scaling law holds for varieties of 
mean fields $h_n$.
From these studies it is seen that the analytical approach by Ershov 
and Potapov \cite{Ershov_Potapov} 
gives estimated values of $F$ which are much smaller than 
those found in the present numerical results.

For the case not satisfying the above condition for $|f'(X)|$,
for instance globally coupled logistic maps, 
the present argument is expected to be qualitatively applicable
under some restrictions.
This is mainly because here, golden values are located in windows, and
the stability of a periodic orbit there causes the so-called
clustering phenomena \cite{Kaneko_GCM} which are out of the present arguments.
On the other hand, collective motion resembling to the present one is also
observed for globally coupled logistic maps,
where the amplitude of this collective motion 
is proportional to the coupling strength \cite{Kaneko3}.
An analytical study \cite{Ershov2} also supports this idea.
These suggests the existence of a certain scaling law of $F$ 
which implies the qualitative applicability of our argument.

Collective motion with a quasi-periodic property is also observed 
in spatially extended systems \cite{Chate_Manneville}.
Since GCM is an idealized model of a long-range coupled system,
it does not take into account anything resembling spatial extension.
We wish to be able to find a way to include the effects of spatial degrees 
of freedom in the description of collective motion.
It is hoped that the discovery presented here will be relevant to
such spatially extended systems also.

%%%%%%%%%%% discussion %%%%%%%%%%%%%

We have concentrated on the amplitude $F$ of the collective motion
without going into its phase-space structure.
The scaling relation discovered for this non-specific quantity holds 
over a wide range of parameter values for which
the type of collective motion changes in various ways.
Based on this point, we suspect that the collective motion 
discussed in the present Paper 
might possess quite unusual properties 
not shared by conventional low-dimensional dynamical systems.
Detailed numerical study of the nature of various types of 
collective motion, which is now in progress, will clarify this point.

%%%%%%%%%%%%%% acknowledgment %%%%%%%%%%%%%%%

The authors are grateful to T.Chawanya for helpful discussion, 
and to Y.Kuramoto, B.Hinrichs and K.Kaneko for encouragement and 
critical reading.
This research is supported by JSPS Research Fellowships.

\begin{figure}[h]
\caption{Two types of collective motion at $a=g_5$ are displayed 
by the return map of $h_n$.
(a) For $K=0.1$, we find a torus representing quasi-periodic motion.
(b) For $K=1.0$, the return map indicates more complicated torus-like motion
possibly accompanied by fine structure.
}
\label{fig:snapshot}
\end{figure}

\begin{figure}[h]
\caption{Amplitude of collective motion $F$ vs $m$ for two values of $K$.
Here $m=-\log_a(2-a)$.
There are several hills and valleys, whose numbers increases as $K$ decreases.
Some golden and silver values, denoted by $g_p$ and $s_p$,
are also displayed for reference.
}
\label{fig:a_F}
\end{figure}

\begin{figure}[h]
\caption{Scaling of $F$ with $K$.
Here $KF$ vs $1/K$ is shown for five golden values $g_p$, $p=3\sim 7$.
The values of $g_p$ are the same as those
in Fig.\protect{\ref{fig:a_F}}.
The data clearly exhibits the linear dependence of $\log(KF)$ on $1/K$.
}
\label{fig:gn0_exp}
\end{figure}

\begin{figure}[h]
\caption{Widths of the hills around a few key golden values.
Here (hill width) vs $1/K$ is shown for five golden values $g_p$ 
($p=3\sim 7$)
whose values are the same as those in Fig.\protect{\ref{fig:gn0_exp}}.
For each golden value, we find a good correspondence 
between the line here and that in Fig.\protect{\ref{fig:gn0_exp}}
}
\label{fig:width}
\end{figure}

\begin{figure}[h]
\caption{
(a) Invariant measure $\rho(X)$ and (b) a snapshot of the distribution 
$\rho_n(X)$ at a certain time step $n$, 
where $a$ is set at the golden value $g_5$.
(a) The invariant measure is given by a $3$-step function.
(b) The instantaneous shape of the distribution for $K=0.1$
is similar to the invariant measure, except that it is accompanied by
peaks with finite widths which appear near each corner of the steps.
The snapshots of $\rho_n(X)$ for all values of $a$ in the neighborhood of 
$g_5$ have a similar property.
}
\label{fig:distribution}
\end{figure}

\begin{figure}[h]
\caption{
Amplitude of collective motion $F$ near a silver value $s_7$ 
is displayed on a logarithmic scale,
where $\Delta a\equiv a-s_7$.
The picture on the left is for values of $a$ smaller than $s_7$,
while that on the right is for $a$ larger than $s_7$.
For smaller $K$, $F\propto |\Delta a|$, so that $F$ decays to zero
as $a\rightarrow s_7$.
On the other hand, $F$ remains finite for larger $K$ 
over the range of $a$ in this figure.
}
\label{fig:s1_exp}
\end{figure}

\begin{figure}[h]
\caption{
$KF$ vs $K^{-1}$ for two golden values 
whose values of $a$ differ only slightly.
The plotting method is the same as that for Fig.\protect{\ref{fig:gn0_exp}}.
The value of $g_4$ is the same as that given 
in Fig.\protect{\ref{fig:a_F}}.
Each golden value is regarded as key golden value
while $K$ is sufficiently small.
However, only the collective motion at one golden value, $g_4$, 
is characterized by the same scaling law in both the small-$K$ and 
large-$K$ regions.
The $K^{-1}$ dependence of $KF$ at the other golden value, $g_{17}$, changes
and this dependence comes to resemble that at $g_4$.
This implies that the hill associated with $g_4$ absorbs 
the hill associated with $g_{17}$.
Thus, the golden value $g4$ continues to be a key golden value,
while $g_{17}$ cease to be such.
}
\label{fig:merge}
\end{figure}

\begin{figure}[h]
\caption{
The point $K_c$ at which the minimum point in the valley 
closest to the key golden value $g_4$ vanishes is investigated.
Here $\Delta a = a-g_4$.
The value of $g_4$ is the same as that given
in Fig.\protect{\ref{fig:a_F}}.
Note that the horizontal scale is logarithmic.
The picture on the left is for the values of $a$ smaller than $g_4$,
while that on the right is for larger values.
When a minimum situated at the silver value disappears at some $K=K_c$, 
the width of the hill associated with $g_4$ becomes expanded.
The values of $K_c$ for those valleys which disappear successively show a 
logarithmic dependence on $\Delta a$.
Although this figure ignores many other valleys which are actually present, 
we believe that the manner in which valleys disappear successively is
consistent with the variation of the hill width with $K$ shown 
in Fig.\protect{\ref{fig:width}}.
}
\label{fig:valley_disappear}
\end{figure}

\begin{figure}[h]
\caption{
A blowup of the phase diagram for the collective motion 
in $m$-$K^{-1}$ space, where $m=-\log_a(2-a)$  (restricted in $2.82<m<3.22$).
The value of the $g_4$ in this figure is the same as that 
in Fig.\protect{\ref{fig:a_F}}.
Each domain filled with black has been determined from the calculation
of the hill width associated with the corresponding golden value.
The golden values represented here are $g_p$ with $p\leq 11$.
The neighboring domains overlap for larger values of $K$.
A precise phase diagram split into an infinite number of domains
could be imagined as an extension of this figure.
}
\label{fig:diagram}
\end{figure}

\end{document}